# Sustainable Online Communities Exhibit Distinct Hierarchical Structures Across Scales of Size

**Authors:** Yaniv Dover[1*], Jacob Goldenberg[2], Daniel Shapira[3]

**Affiliations:**

[1] Jerusalem School of Business Administration, Hebrew University, Jerusalem, Israel 91905.

[2] Arison School of Business, Interdisciplinary Center Herzliya, Herzliya, Israel 4610101; and a Visiting Professor at the Graduate School of Business, Columbia University, New York, New York 10027.

[3] Guilford Glazer Faculty of Business and Management, Ben-Gurion University, Beer Sheva, Israel 84105.

*Correspondence to: yaniv.dover@mail.huji.ac.il.

**Abstract**:

**Online communities exist in many forms and sizes, and are a source of considerable influence for individuals and organizations (*1, 2*). Yet, there is limited insight into why some online communities are sustainable, while others cease to exist *(3)*. We find that communities that fail to maintain a typical hierarchical social structure which balances cohesiveness across size scales do not survive, and can be distinguished from communities that exhibit such balance and prevail in the long term. Moreover, in an analysis of 10,122 real-life online communities with a total of 134,747 members over a period of more than a decade, we find that mapping the community social circle structure in the first 30 days of its lifetime is sufficient to forecast the survival of the community up to ten years in the future. By varying calibration time frames, the aspects of the social structure that allows for predictive power emerge and fixate within the first couple of months in a community's lifetime.**

A community can exhibit a variety of social structure configurations. However, in actuality, not all hypothetical structures may be observed: Only those configurations that promote survival at each scale are likely to be observed, while other communities with sub-optimal structure configurations are expected to fade away. Indeed, in our data, we find that communities that thrive are those that exhibit distinct structural network features that are scalable across sizes: We identify a "valley of stability" that marks a range of structures for which communities preserve sustainability. These findings suggest an analogy to the stability of physical matter: Only structures that exist in an optimal balance between contradictory forces across scales of size survive for relatively long periods.

While the hierarchical nature of social communities is well documented [e.g., (*4-9*)], its association with network stability has yet to be established. This hierarchy captures valuable information on the stability of networks and can indicate the "health" level of that network and be an important predictor of stability.

Individuals gain benefits from group membership (*10*), a major benefit being the common relevance of group members to one another and the consequent resource exchange within the group. An important factor in the stability of a social group is its level of cohesiveness (*11*). A well-functioning cohesive group supplies information, security, and a variety of other important resources to its members (*12, 13*). In addition, previous studies have shown that access to outside connections and social circles leads to benefits and information that are associated with the outside world [e.g., finding a job through weak ties (*15*), or exposure to innovation (*9, 16*)]. In general, investment in a few links with non-kin "friends" that connect unrelated families can potentially increase efficiency for information exchange (17). Obviously, each member of a group would want to benefit from both global and local connections, but there is a basic constraint: group members have limited time and effort to spend on maintaining relations. This leads to a natural tradeoff between connection choices. Community stability is, therefore, a consequence of a subtle balance between local (direct) and global (more exploratory) benefits. The members of each existing small clique may decide to spend their resources to interact strictly with other members of their clique. This allows them to benefit from their group's resources and protection, but increases the risk of fragmentation of the clique from the rest of the community. Alternatively, they may choose to focus on relationships outside the clique to seek new opportunities and exposure to new ideas, and to increase their scope. This, in turn, may result in risking their own clique's long-term stability. A compromise between these competing sets of needs would be a mix of the two approaches: group members connect to larger external circles, but only to those that their own clique peers also connect to. Consequently, cliques are able to maintain their cohesive structure, while at the same time to overlap with other cliques, simultaneously conserving the cohesiveness of the whole community. The result is a hierarchical structure in which the extended social circles are an aggregation of smaller scale circles, but integrity is conserved for both scales. The generalization to higher scales is straightforward. For each scale, members of the social circle of that scale will balance between relationships within or outside it.

In fact, we maintain that unlike common approaches which rely on a dichotomy between the individual- and aggregate-level structure—identifying and using social sub-communities across varying size scales as the units of analysis is a much more useful and constructive approach. This type of multi-scale cohesiveness facilitates long-term sustainability for the community. A mapping of the profile of the cohesiveness across scales is, therefore, a measure of the community's health in terms of social stability. In order to examine this conjecture, we developed a mapping approach to detect the social circles across scales and then, from the mapping, calculated a cohesiveness-vs.-scales profile. We defined a social circle's cohesiveness as the overlap of interests among its members and operationalized it by using the clustering coefficient as the measure of interest overlap (*18*). Then we tested whether the cohesiveness-vs.-scales profile is a good diagnostic of community stability and evaluated its predictive power.

We used a multi-scale approach to map social circles (i.e., sub networks) per given scale level of cohesiveness. Let $M_S$ be the social circles' mapping of a social network of $N$ individuals for a given scale $S$, where $S$ is a positive integer number representing the radius of a cohesiveness sphere. The mapping $M_S = \{C_{\alpha,S}\}$ is the set of social circles $\{C_{\alpha,S}\}$ for a given $S$ such that $C_{\alpha,S} \in M_S$ if the two following conditions are satisfied:

(i) For each and every pair of individuals *i* and *j* in the network that belong to social circle $C_{\alpha,S}$, the shortest path $d_{i,j}^{C_{\alpha,S}}$ between them within the social circle $C_{\alpha,S}$ satisfies: $d_{i,j}^{C_{\alpha,S}} \leq S$.

(ii) There is no inclusion relationship between social circle $C_{\alpha,S}$ and $C_{\beta,S}$ of mapping $M_S$ for $\forall \alpha, \beta$, and $\alpha \neq \beta$. Namely, for any two social circles $C_{\alpha,S} \in M_S$ and $C_{\beta,S} \in M_S$ for which $\alpha \neq \beta$, the condition is $C_{\alpha,S} \not\subset C_{\beta,S}$ and $C_{\beta,S} \not\subset C_{\alpha,S}$.

Figure 1 illustrates a simple example of a mapping for a hypothetical network with ten nodes. In the figure, it is possible to see how small circles of high cohesiveness merge into larger ones of lower cohesiveness. For instance, the tightest cohesiveness sphere $S = 1$, condition (i), results in the mapping $M_1$, which is the mapping of the fully connected groups in the community, i.e., the network cliques (4). In the figure, it is possible to see that $M_1$ includes three high-cohesiveness circles, or cliques. Condition (ii) dictates that each separate circle is not fully contained within another clique although they can still partially overlap. If the basic a-priori units of analysis of the network are the nodes, the clique map ($M_1$) is the mapping of the one-scale-higher "natural" components of the network (S=1 in Fig. 1). Next, $M_2$ is the mapping of the second-order cliques (S=2 in Fig. 1), and so on. The last stage includes only a single social circle—the entire community.

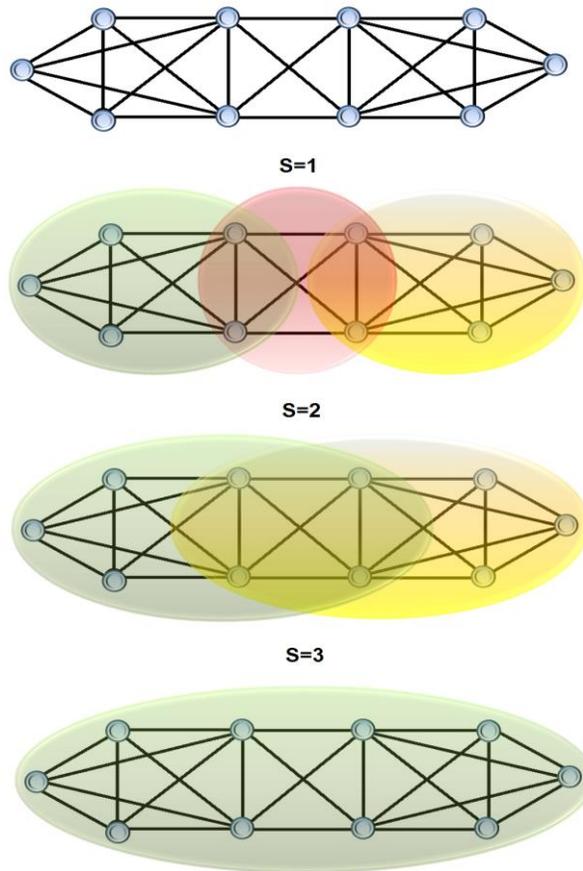

**Fig. 1**. An example of social circles mapping for an illustrative ten-node network. The network itself, with no mapping, is shown at the top of the figure. The mapping level, *S*, is given above each corresponding mapping. The full explanation of how the social circles mapping was calculated is given in the supplementary information, S1.

Figure 2 illustrates the mapping process for a real-life sample of 43 community members in a community that was created to host discussions of users interested in a contemporary religious lifestyle. Figure 2A shows a plot of the network of interactions between the members of that community. The red area marks the one and only detected social circle at the level of $S = 4$, i.e., the whole community. In practice, the figure shows that the whole community is connected at a distance of either four connections or less. The one-step lower scale $S = 3$ (Fig. 2B) shows a more granular picture of the structure, in which the average social circle size at that scale is around the size of a characteristic band (*6*). The social circles at this scale seem to be cohesive and practically woven into one another, a positive sign of stability. Some of the most popular topics in each separate social circle are given in Figure 2A–D. A post factum inspection of the figures and text reveals that the social circles mapping closely corresponds to topical mapping. Users form small groups of unique common denominators, which join into larger groups of more widespread common denominators. This happens across scales. The detailed analysis of the major topics within each detected social circle is described in Section S2.

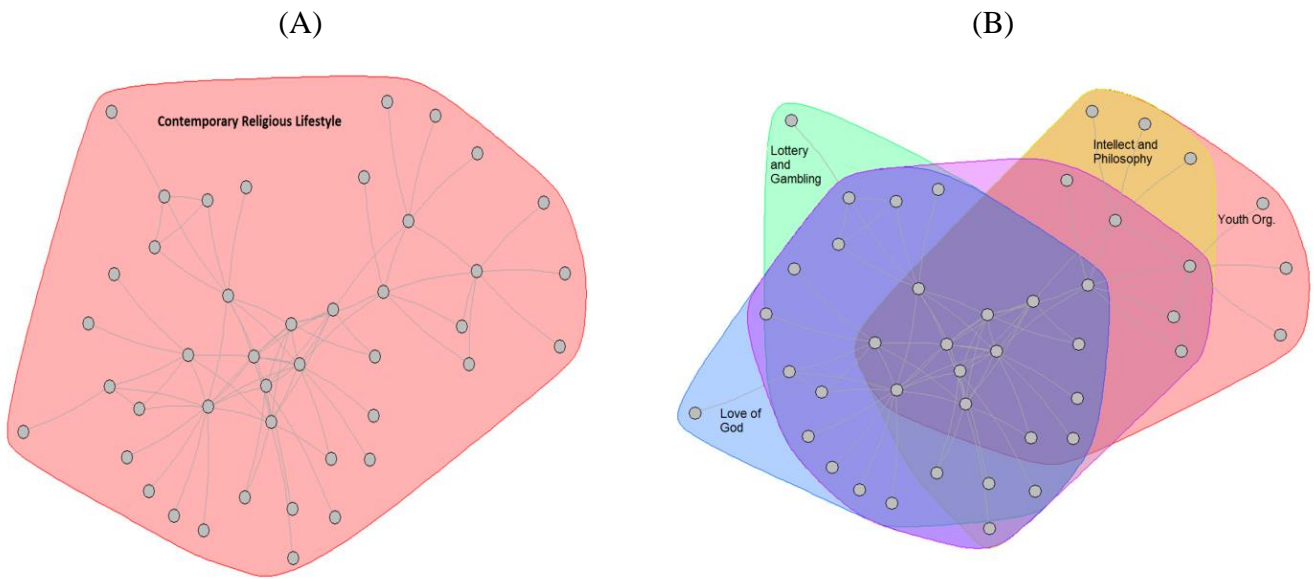

(A) (B)

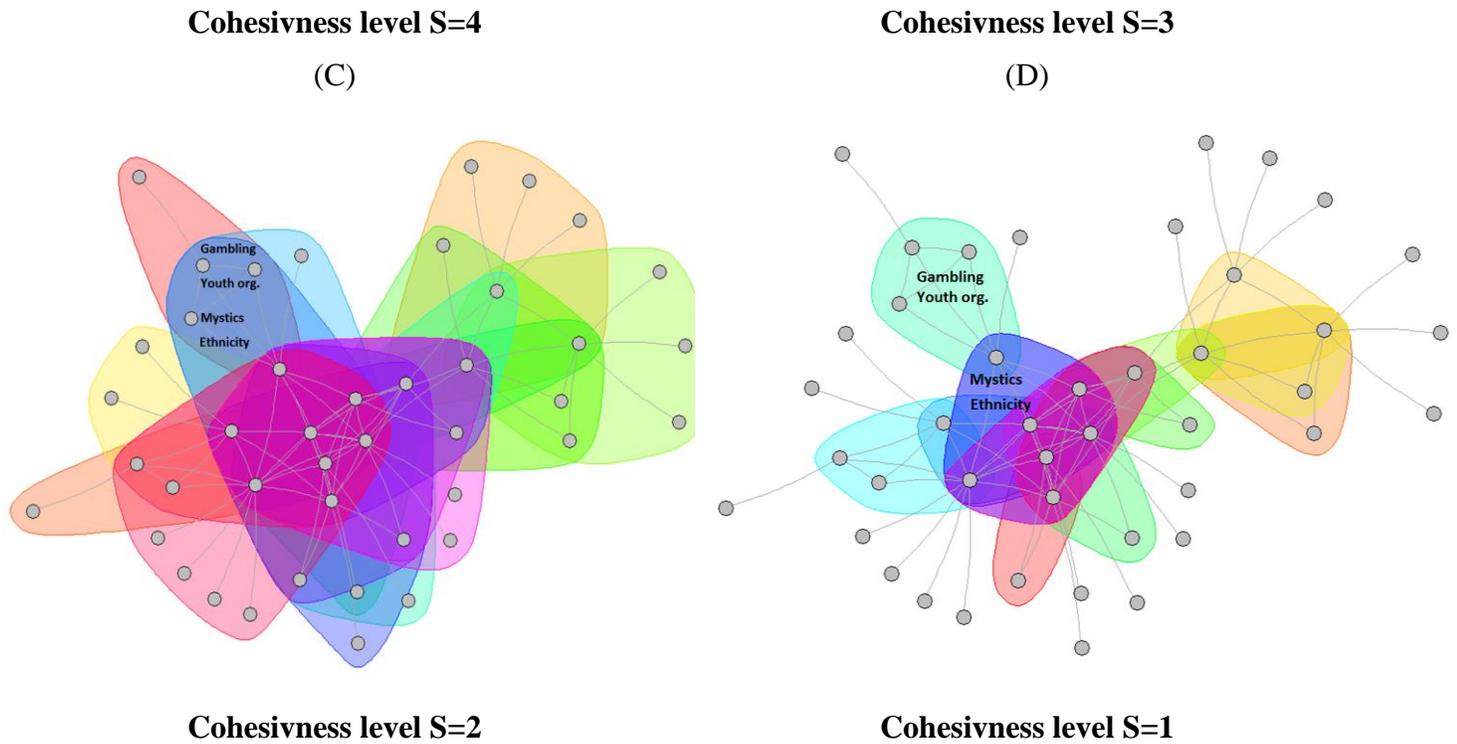

**Cohesivness level S=2**  **Cohesivness level S=1**

**Fig. 2**. A schematic illustration of a contemporary religious lifestyle community's network and social circles per each scale: (A) S=1, (B) S=2, (C) S=3, and (D) S=4. Social circles appear in distinct colors to show differentiation.

In order to translate the complex multi-scale structure mapping into a usable community sustainability measure, we developed a cohesiveness-vs.-scales profile, a measure of how finer scale circles aggregate to become larger ones across scales. As aforementioned, we expected a stable community to exhibit a characteristic cohesiveness-vs.-scales profile that is the result of the balance between local and global considerations across scales. The social circle mapping process was done using expanding spheres of path lengths to identify scales. This allows a straightforward means of defining the discrete scales (i.e., corresponding to the maximal discrete path length within a circle). But, in order to construct a cohesiveness-vs.-scales profile with higher resolution, we used a continuous measure, namely, the clustering coefficient of the social circles. Figure 3A presents the mean of the cohesiveness-scale profiles across all communities in the data. In the figure, for illustrative purposes, we marked the group sizes with the labels taken from Zhou et al. (6). Interestingly, the characteristic profile in Figure 3A seems to curve sharply up for groups of 7–8 and above (sympathy group scale). This suggests that in order for a group, in this context, to grow above this number of members, it must conserve cohesiveness. Furthermore, for groups of 200 members (clans), mean cohesiveness does not fall below 55–60%, suggesting that this may be a minimum stability requirement for such large groups. Figure 3B contrasts the profile for communities with lifetime above (blue) and below the median (red). Interestingly, it seems that cohesiveness of up to ~8–10 members is not associated with stability, while the curve shows that cohesiveness above this group size is a determining correlate of it.

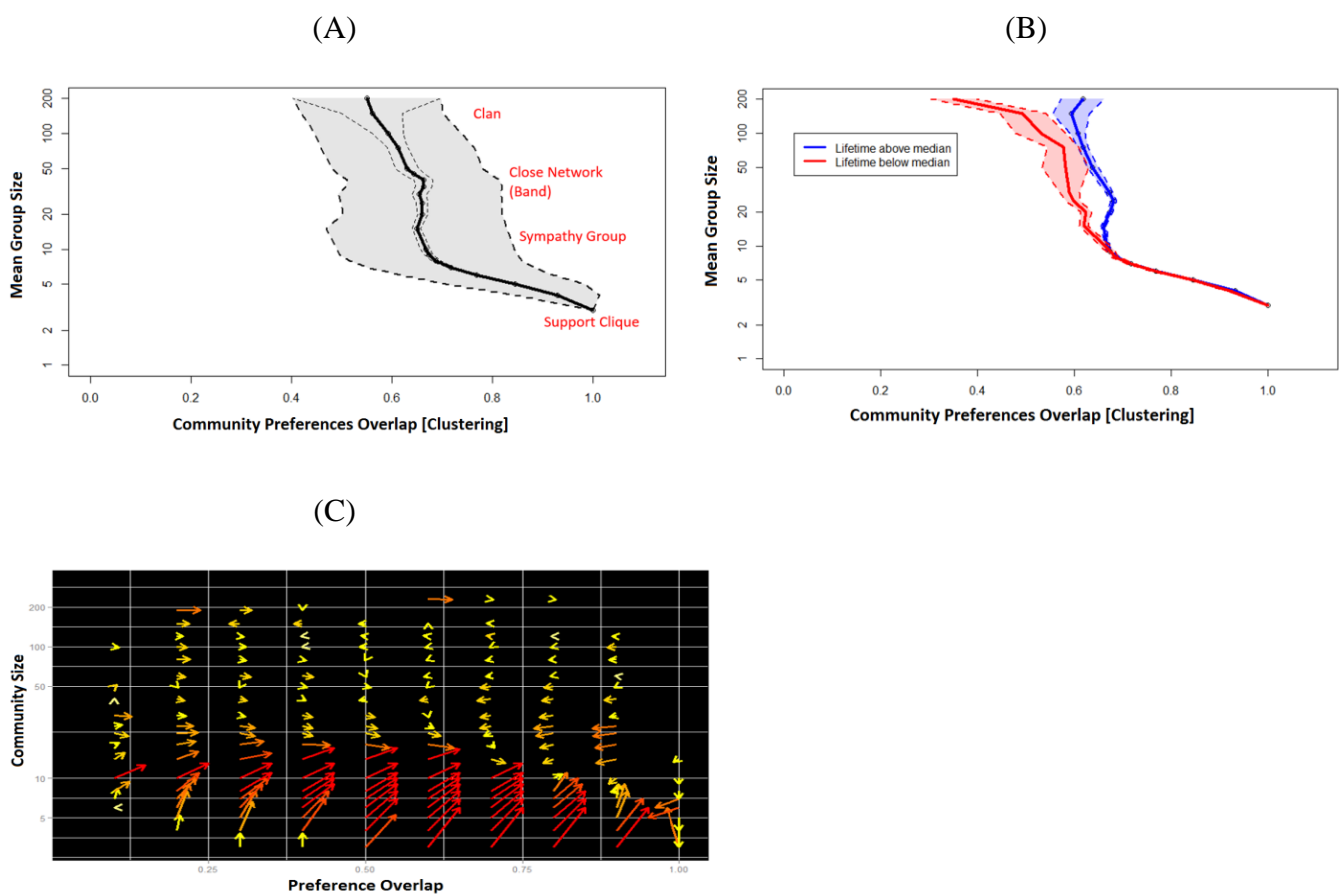

**Fig. 3**. Cohesiveness-vs.-scales profiles. The x-axis measures the mean clustering within a social circle, and the y-axis shows the mean size of a social circle with given clustering. (A) The mean profile across all

10,122 communities. The thick (thin) dashed lines and the grayed area denote the standard deviation (95% confidence intervals) of the cohesiveness-scale profiles. (B) Profiles for communities with lifetime above (blue) and below (red) the median. (C) A vector field graph illustrating the tendency of the cohesiveness-vs.-scales curve to change over time, calculated using all communities for monthly time periods.

To further uncover the underlying "pressures" on the cohesiveness-scale profile, Figure 3C illustrates its dynamic vector field. An arrow in the figure represents the tendency of a curve of a community to change over time (*19*). For example, in the figure, it is visible that small communities can hardly exist in low-cohesiveness contexts. This can be seen by the long arrows in the lower half of the graph. The arrows point upwards and to the right, meaning that a small community with low cohesiveness strongly tends to become more cohesive if it is to grow. The valley of stability, i.e., the region in which the pressure is close to zero, roughly corresponds to the mean curve shown in Figure 3A.

Finally, we tested the predictive power of the cohesiveness-vs.-scales profile: We divided communities in the data into four tiers of size, corresponding to the four size quartiles. Then, we compared the performance of the model based on the cohesiveness-vs.-scales approach to the performance of a naïve benchmark model that simply uses the early-stage size of the community for prediction. Both models were tested to see how well they could predict a community's size at a future time by using data only from the first 30 days of the community's existence. We used the holdout sample approach in which we calibrated a statistical model (*20*) on a randomly chosen half of the communities and tested the model's prediction on the remaining half. We did this for a large sample of randomized sample choices (1,000) and reported the average rate of prediction success in Table 1 (for the full details and robustness checks, see SI Appendix, Section 5).

| Horizon of prediction | Success rates: Benchmark model | Success rates: Social circles approach | Difference between rates: t test value (p value) |
| --- | --- | --- | --- |
| 120 days | 36.6% | 43.5% | 26.8 (<2.2e-16) |
| 1 year | 31.8% | 42.9% | 85.4 (<2.2e-16) |
| 5 years | 25.1% | 39.8% | 105.3 (<2.2e-16) |
| 10 years | 25.1% | 38.6% | 110.7 (<2.2e-16) |

**Table 1**. Prediction success rates as a function of prediction horizon for the benchmark and social circle model. The calibration period was the first 30 days of the community's existence. The far right column presents the t value of a Welch two-sample t-test of the statistical difference between the success rate of the social circles model and the benchmark model.

The results for varying prediction horizons are presented in Table 1. The table shows that the social circles approach is statistically significantly superior to the benchmark model for all prediction horizons. Notably, for the long-term prediction horizon, the benchmark model loses almost all of its predictive power and is reduced to a simple coin tossing (i.e., 25%). Conversely, the social circles model retains its predictive power (*21*) and exhibits a high level (38%) of success even though the prediction horizon is much longer than the calibration period. Figure 4 shows the prediction success rate of the social circles model with a ten-year horizon, but for varying calibration windows ranging from a two-week window to a four-month window. The fact that that prediction power increases as the calibration window becomes longer is a validation of the predictive usefulness of our suggested approach. Furthermore, it is interesting to note that prediction saturates at a period of around two months. This suggests that this period is the characteristic time frame required for the hierarchical social circles structure to stabilize.

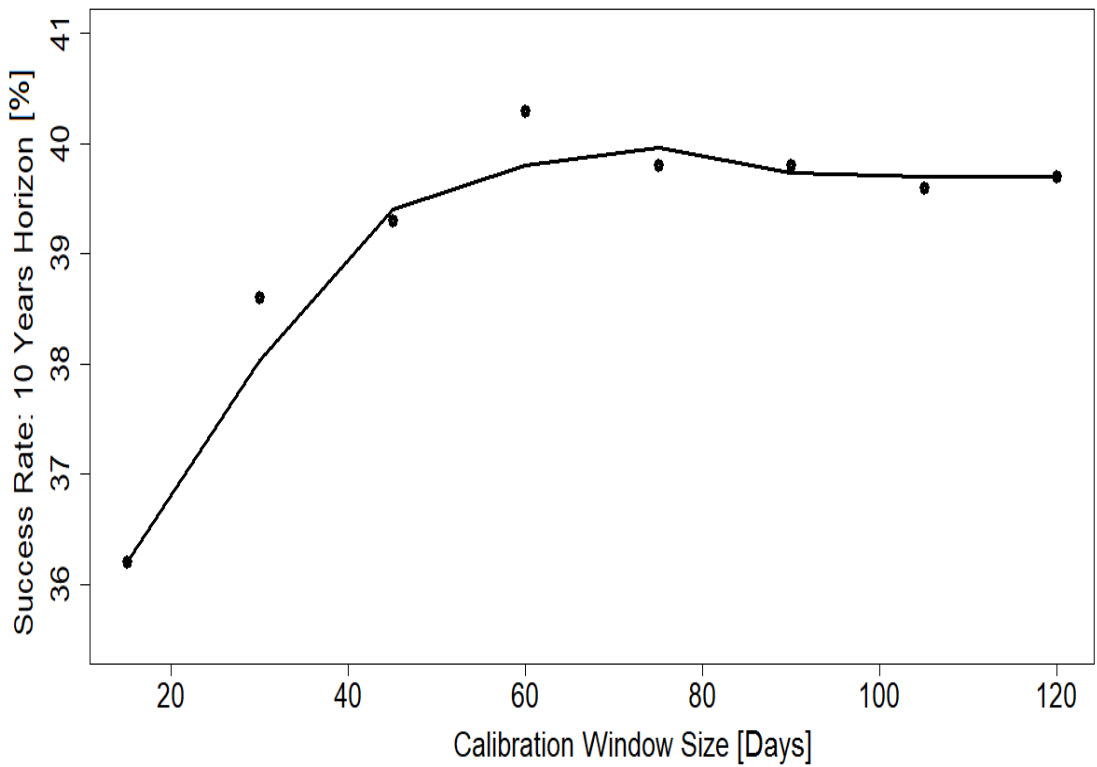

**Fig. 4**. Prediction success rates as a function of the calibration window. The prediction horizon used to calculate this curve is ten years. The points denote the prediction rates, and the solid line denotes the smoothed curve. The prediction curve of the benchmark model is not illustrated as it was mostly independent on a calibration window at a level of ~25% success rate.

The prediction study shows that the mapping of social circles across scales is informative and is an important indicator of the integrity of a community. In sum, our approach seems to be useful both in terms of insights gained into the underlying hierarchical structure of a community, and as a way to gauge the health and stability of a community in the present and future. Our findings suggest that it is important to understand and capture the hierarchical nature of the network structure when trying to model online communities.

**Methods:**

**Data collection and preparation**

The data were collected from an Israeli-based forum discussion website called Tapuz. The website includes company-created forums which constitute the main feature of the site. The decision to start or close these forums is made solely by the company that operates the platform. On the website, a satellite feature exists called "communes." These are effectively theme-based discussion forums that originate and are controlled by the site's users, i.e., an opportunity to investigate spontaneous emergence and collapse of online communities. Each user-controlled community has a stated theme (e.g., a community of discussions of news) and a discussion page that is unique to it. The interface of the discussions within each community allows users to post direct answers to other messages, such that the message's sender and recipient are known. We randomly collected 10,122 such communities with a total of 134,747 members over a period of over a decade (about 12 years). In this paper, we define a community to be one distinct user-controlled forum (i.e., "commune"). To measure the network of social interactions for community k within a time slice $\Delta t$, we linked between two people *i, j* if person *i* posted a message directly to person *j* within the page of community *k*, within time slice $\Delta t$.

**The cohesiveness-vs.-scale profiles extraction**

The multi-scale network mapping approach is based on social circles mapping per given discrete level of cohesiveness, as defined in the above conditions (i) and (ii). Next, we describe how we constructed a continuous profile of cohesiveness across scales from multi-scale social circles mapping. The result of the social circles mapping algorithm across scales is the mapping $M_S = \{C_{\alpha,S}\}$ for $S = 1,2,..,S_{max}$. Practically, for each scale $S$, we have a set of social circles $C_{\alpha,S}$ for which $\alpha$ is the index of the circle within the mapping of $S$. In order to develop an indicator of cohesiveness across scales, we conducted the following steps:

1. We calculated the (continuous) level of cohesiveness of each social circle ($C_{\alpha,S}$) by calculating its average clustering coefficient. In our application, clustering served as a proxy for cohesiveness.
2. We defined scale (social circle size) bins.
3. Per each size bin, we collected all social circles within the community within the range of sizes in the bin. We then calculated the mean cohesiveness of the social circles within that bin.
4. We plotted the mean cohesiveness (x axis) versus size bin (y axis).

The result of the above procedure is a curve of cohesiveness as a function of size. This curve characterizes how cohesiveness changes across increasing scales. An example for such a profile, calculated for a contemporary religious lifestyle community, is given in Figures 3A and 3B.

**Generation of the cohesiveness-vs.-scales vector field**

The cohesiveness-vs.-scale curve measures the way that social circles of varying scales aggregate. It is interesting to look into the "pressures" that exist for this curve over the lifetime of a community. In order to do this, we calculated and plotted a vector field of the time dynamics that exist within the cohesiveness-vs.-scale space. We did this in the following way:

1. For each community, the cohesiveness-vs.-scale (CvS) profile was calculated for time slices of 30 days each.
2. Next, we calculated the tendency of points on the CvS curve to change over time. This was done by comparing the CvS of a community at time t to the CvS at t-1.
3. A requirement is to pinpoint which point of the CvS in t corresponds to which other point of the CvS in t-1, so that the movement of each point in the respective CvS curve can be calculated. To do this, we used the minimal Euclidean distance in the cohesiveness–size space between points on the CvS curve at time t-1 to points in the CvS curve at time t.

4. Per each point at t-1, the point closest to it at t was chosen. We then calculated the magnitude of the movement of the curve for that point.

The result of the above procedure is that per each point (binned) in the cohesiveness–scale space, there was a corresponding set of vectors. Each vector in the plot was a measure of the tendency of each point on the CvS curve to move over time. To visualize that vector field, the mean vector at each point was plotted and colored according to its magnitude (e.g., Figure 3C). The vector field then illustrates the stress field on the CvS curve, over time, of an average community in the data.

**Details of the detected topical structure of the contemporary religious lifestyle community example**

The online communities in our dataset are based on stated discussion themes. This opened an opportunity to gauge the interests of the members of each detected social circle outside the focal community. This mapping of the themes for each social circle can indicate the common denominators that drive the multi-scale structure of an online community. An example for a comparison between the multi-scale structure and its thematic mapping is given here.

To generate the thematic mapping per each social circle, we measured the frequency of the activity of the members of that circle in *other* forums of other topics. In other words, we counted the frequency in which an external topic/forum was engaged in by the members of a given social circle. We then, per social circle, ranked the frequency of the appearance of the topics in order of decreasing popularity. We chose the top five themes that were popular within the circle as the characteristic topics of that circle, and used this information to gauge the differences between common themes of the social circles.

The full details of the example we investigated here, i.e., the contemporary religious lifestyle online community, is given in the supplementary materials.

**Details of the prediction study**

The goal of the prediction study was to show that the cohesiveness-vs.-scale (CvS) curve provides useful information about the stability of the group and, therefore, provides prediction power of the long-term size of a community.

In order to test this, we constructed a statistics prediction model that uses only the initial 30 days of activity in a community to predict the long-term size of that community. The size of the community in the future horizon, i.e., the number of active members, was measured only using the activity occurring *after* the initial 30-day calibration period. To gauge the benefit of using the CvS, we compared its predictive capacity to a baseline model that only included the early-stage size of the community and the average density of links in the community, expressed by the mean degree, in the initial 30 days. The prediction horizons were chosen to be: 60 days, 120 days, one year, five years, and ten years. We compared the success of the prediction of the base model to one based on the CvS model. We used three types of models as the base of the prediction method: a Poisson count regression model and two models that allow over-dispersion, a negative binomial model and a quasi-Poisson model. The results for all models agreed qualitatively. The details of the model and results are given in S3 in the supplemental materials. We also examined the prediction success rate at a ten-year horizon for various calibration times starting from 14 days to 120 days. These results are presented in Figure 4.

18. The overlap of preferences within a social circle can be obtained by using richer data variables, if they exist and can be measured.

19. See explanation of the method in the Methods section.

20. In order to test our power of prediction of community size (i.e., a count variable), we used a Poisson model, a Negative Binomial model, and a Quasi-Poisson model.

21. The prediction study shows that the average community in the long-term retains the "memory" of the hierarchical social circles structure that already exists in the early stages of the life of the community.

# Supplementary Materials

## Sustainable Online Communities Exhibit Distinct Hierarchical Structures Across Scales of Size

Yaniv Dover, Jacob Goldenberg, Daniel Shapira

# S1 An illustrative example of social circles mapping for two simple networks

Figure S1 illustrates an example of the social circles mapping process for two simple networks of slightly different structure (Figure S1a and S1b). Its purpose is to demonstrate the mapping process for a simple example and to show how this method can provide scale-specific insight into the hierarchical structure of networks. Each row in the figure corresponds to the mapping of a certain level of cohesiveness $S$. For each mapping, the detected social circles are marked in the figure by ellipses of different colors. The mapping in the second row is for the bottom scale of the direct friends level ($S=1$) and reaches the top-most scale of $S=3$, in which all members of the community can be considered to be in one social circle. Specifically, beginning with the second row, even though the difference between the two networks is only two links, the mapping $M_1$ shows a distinct difference in social circles between the two networks. The left column network (Figure S1c) shows a tighter construction between cliques relative to the right column network (Figure S1d). The social circles of $M_1$ for the left column network overlap and are practically woven into each other. In contrast, the circles in Figure 1d do not overlap and, assumingly, are at higher risk of separation.

The mapping of the next sphere of cohesiveness $S=2$ (Figure S1e and S1f) is the mapping ($M_2$) for social circles in which all members are either direct friends or at most friends of friends. The cohesiveness within each social circle is, therefore, lower in this case, and so the characteristic size of a social circle is larger, as can be seen from the figures. Interestingly, at this scale, there is no major visible difference in the cohesiveness structure between the two networks: both are composed of two same-sized overlapping social circles. The algorithm then stops for $S=3$, illustrated in the fourth row, in which the whole community is one big social circle for both networks.

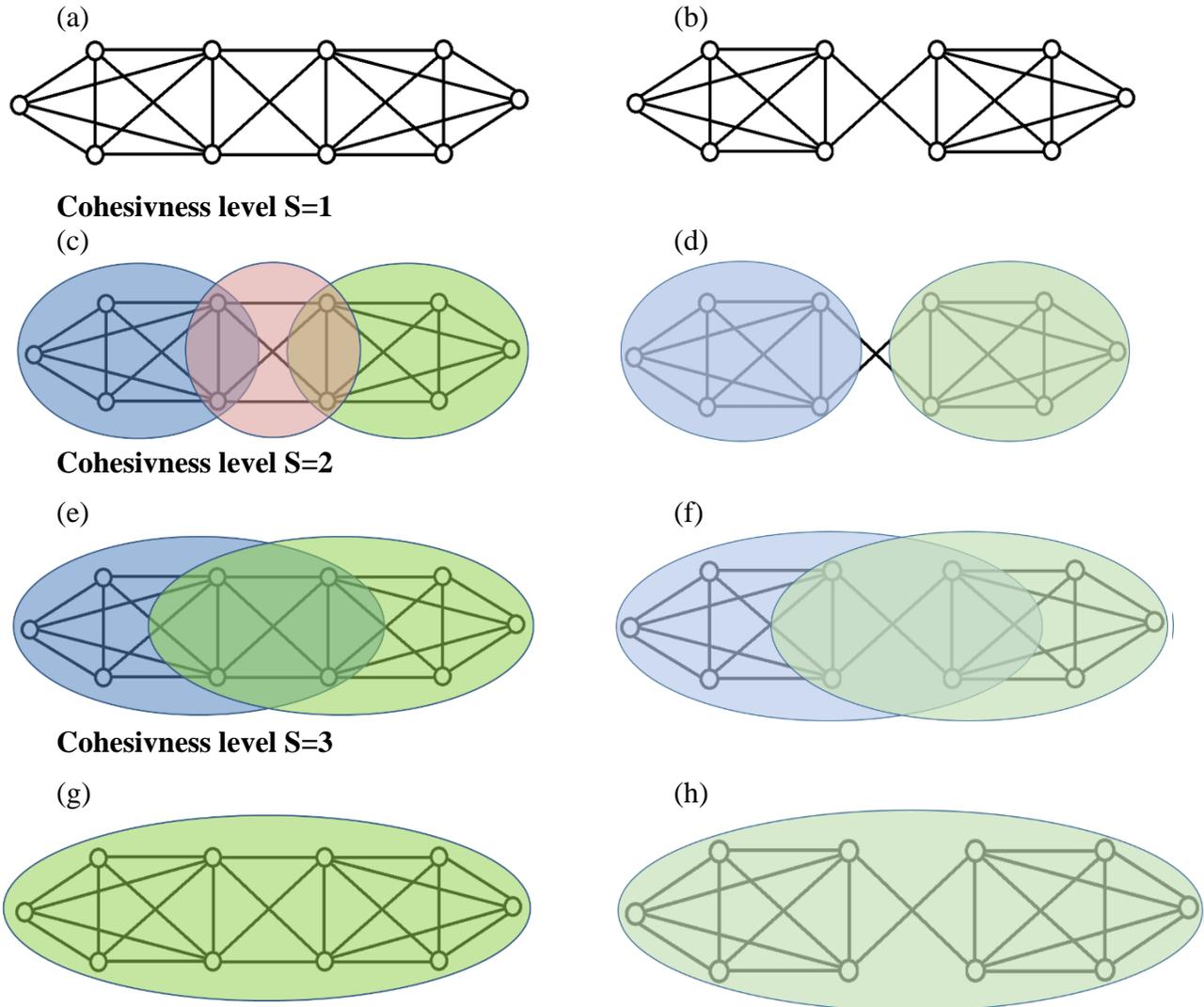

Figure S1: An illustration of the social circles multi-scale mapping process for two slightly different networks. In the first row (S1a and S1b), the structure of the two networks is shown. The second row (S1c and S1d) shows the detected social circles for S=1. The third row (S1e and S1f) shows the mapping for S=2, and the fourth row (S1g and S1h) shows the mapping for the top-most scale S=3.

This basic example shows that the difference of two specific links removed between networks will only have an effect on the $S = 1$ scale. The rest of the scales are unaffected. This demonstrates that social circles mapping is a way to decompose the hierarchical social network structure into the social structures that exist in each separate scale. In S3, we show how to convert the multi-scale social circles mapping into a profile of the cohesiveness-vs.-scale of the network.

# S2	Explanation of the detected topical structure of a contemporary religious lifestyle community example

The online communities in our dataset are based on stated discussion themes. This opens an opportunity to gauge the interests of the members of each detected social circle outside the focal community. This mapping of the themes for each circle can indicate the common denominators that drive the multi-scale structure of an online community. An example for a comparison between the multi-scale structure and its thematic mapping is given here.

To generate the thematic mapping, per each social circle, we measured the frequency of the activity of the members of that circle in *other* forums of other topics. In other words, we counted the frequency in which a topic/forum was engaged in by the members of a given social circle. We then, per social circle, ranked the frequency of the appearance of the topics in order of decreasing popularity. We chose the top five themes that were popular within each circle as the characteristic topics of that circle, and used this information to gauge the differences between common themes of the social circles.

The example we investigate here is for the contemporary religious lifestyle online community discussed in our paper. The top level, or scale, of the community is given in Figure 1a, which includes the whole community at S=4. The most interesting scale is the S=3 level (Figure S1b), as it shows the crudest fragmentation of social circles in the community. Note that each social circle in this level is about the size of a band [as in (6)], which is in the order of magnitude of several dozen members. Table S1 lists the top themes per social circle shown in Figure S1b. The group's color, as it appears here, is given in parentheses.

Table S1: List of the top five themes for each social circle as shown in Figure 1b in the main paper

| Popularity rank | Group 1 (pink) | Group 2 (yellow) | Group 3 (green) | Group 4 (blue) | Group 5 (purple) |
|---|---|---|---|---|---|
| 1 | Contemporary Religious Lifestyle | Contemporary Religious Lifestyle | Contemporary Religious Lifestyle | Love of God | Contemporary Religious Lifestyle |
| 2 | Computers and Cellphones for the Religious | Computers and Cellphones for the Religious | Computers and Cellphones for the Religious | Contemporary Religious Lifestyle | Love of God |
| 3 | Religion | Religion | Religion | Computers and Cellphones for the Religious | Computers and Cellphones for the Religious |
| 4 | News for the Religious | News for the Religious | News for the Religious | Religious Themes | Religious Themes |
| 5 | Religious Youth Organizations | Intellect and Philosophy for the Religious | Gambling and Lottery for the Religious | News for the Religious | News for the Religious |

The topics are ordered by popularity, as explained in the text, from top to bottom. Topics that exhibit the difference of each social circle from other social circles are marked in yellow. It is important to reiterate that the social circles themselves are defined via the structural method and not by the thematic contents. Therefore, the association between the structural crystallization and thematic content is informative, and evidence that the structural analysis is useful.

In general, Table S1 shows the high thematic overlap that exists between all S=3 social circles, consistent with the high overlap seen in the network structure. Each circle differs though, mostly by one main topic of interest. For example, the unique theme for Group 1 is youth organizations. Group 2, on the other hand, is composed of members who are interested in philosophy and intellect. Groups 4 and 5 differ in the popularity of the "Love of God" theme. For group 4, "Love of God" is the most popular, but for group 5, it is second to contemporary lifestyle. In other words, Table S1 shows not only the prevalent common denominators in the community, but also the more intricate structure of sub-themes.

If we delve into the higher resolution of the S=2 structure (Figure 1c in the main paper), we can investigate other examples of the thematic structure behind the social structure. The members of the blue social circle in Figure 1c seem to be a union of people who are interested in gambling, a specific youth organization, mystics, and subjects related to ethnicity (see labels in the figure itself). To understand the inner structure of that social circle, we delve deeper to look at the S=1 mapping (Figure 1d). The topic analysis shows that, indeed, the blue social circle in Figure 1c breaks down into two tighter $S=1$ circles. Members of the cyan social circle in Figure 1c seem to be mainly interested in gambling and youth organizations, while members of the purple circle focus on themes related to mystics and ethnicity. Again, this is an example of how multi-scale mapping can provide insight into the mutli-scale drivers and motivations behind community member participation.

Interestingly, at the highest resolution level $S=1$, Figure 1d also shows that the social circles structure is not of very high cohesiveness. The yellow and orange social circles on the right-hand side of the figure seem to be loosely connected to the rest of the circles, only through a single person. This suggests a possible weak link in the structure with negative implications for stability. In the following section, S3, we show how we develop an indicator that quantifies cohesiveness across scales with which it is possible to monitor and analyze the "health" of the hierarchical structure. With this indicator, we show that it is possible to gain insight into stability and prediction power.

## S3    Details of the prediction study

The goal of the prediction study was to show that the cohesiveness-vs.-scale (CvS) curve provides more information about the stability of the group and, thus, supplies useful prediction power for the long-term size of a community.

In order to test this hypothesis, we constructed a statistics prediction model that uses only the initial 30 days of activity in a community to predict the long-term size of that community. The size of the community in the future horizon, i.e., the number of active members, is measured only using the activity done after the initial 30-day calibration period. To gauge the benefit of using the CvS, we compared its predictive capacity to a baseline model that only included the early-stage size of the community and the average density of links in the community, expressed by the mean degree, in the initial 30 days. The prediction horizons were chosen to be: 60, 120 days, one year, 5 years, and 10 years. We compared the success of the prediction of the base model to one based on the CvS model. Further, we also compared the results to another more intricate benchmark model that included the size of the community and the average degree in the community in the first 30 days. The results do not differ qualitatively, as Tables S2 and S3 show. We used three types of statistics models in the base of the prediction method: a Poisson count regression model and two models that allow over-dispersion, a negative binomial model and a quasi-Poisson model. All models agree qualitatively. We present the results of the quasi-Poisson model. In terms of data, out of 10,122 networks in the set, only the activity over 3,059 networks was high enough in the first 30 days to allow a measurement of the CvS for that early period.

**Specification**: The basic form of the benchmark model for the prediction is:

$$Log(E(N^{longterm} \mid N^{30days}, D^{30days})) = \beta_0^0 + \beta_1^0 \cdot N^{30days} + \beta_2^0 \cdot D^{30days} \quad\quad (S1)$$

In equation S1, $N^{longterm}$ is the size of the community in the long term, i.e., the dependent variable. The predictors $N^{30days}$ and $D^{30days}$ are the size of the community and the average degree in the initial 30-day period, respectively. Namely, we aim to predict the order of magnitude of the community size in the long term, given the independent variables provided in the initial period of calibration [i.e., $Log(E(N^{longterm} \mid \text{Indep. Vars.}))$]. To extract predictors from the CvS curve for the CvS-based model, we use 10 bins of cohesiveness to span the range of cohesiveness of 0 to 1. Each predictor used in the model was prepared using one of the 10 bins of the CvS curve. This, in effect, is the way we chose to model the shape of the CvS curve itself in the regressions, as broken into 10 predictors. For example, to calculate the predictor for cohesiveness of 0.9 (i.e., the 9[th] predictor), we used the value of the initial 30-day CvS curve for cohesiveness within the 0.9 bin (i.e., the characteristic social circle size at the cohesiveness value of 0.9). In the case where at a certain level of cohesiveness the characteristic social circle was in fact the entire community, the values of the CvS curve for the cohesiveness levels that were below this level were defined as the entire community size as well. Furthermore, we expect each point in the CvS curve to be an attractor. A value for, e.g., predictor 9 that is too high would be detrimental to the stability of the community, just as a value that is too low. This implies a need to use non-linear modeling of the predictors approximating the CvS curve. Therefore, we chose the functional dependence of the predicted community size to be a second order polynomial, a parabola. The CvS model is then:

$$Log(E(N^{longterm} | N^{30days}, C_i^{30days})) = \beta_0 + \beta_2 \cdot D^{30days} + \sum_{i=1}^{10} \left[\beta_3 \cdot C_i^{30days} + \beta_4 \cdot (C_i^{30days})^2\right] \quad (S2)$$

Where $C_i^{30days}$ is the CvS value per bin, which appears both as a linear term and a squared term in S2.

**Estimation**: To test prediction, we divided the 3,059 communities into two randomly assigned groups: the calibration sample and the hold-out sample. The process was two-staged and was done separately per each length of future horizon. The stages were: (1) to estimate the model on the calibration sample and (2) to predict the size of the community in the long term. Then we tested the rate of success of the prediction in the hold-out sample for both the benchmark and CvS models. To estimate the rate of prediction accuracy, we divided the communities into four quartiles of community size (i.e., four groups of equal count). The performance metric was the rate of success that a given model was able to predict the correct community size tier in a given future period. By comparison, a simple non-informative (uniform four-sided) "coin toss" would be successful, on average, 25% of the time.

To rule out any flukes stemming from a specific choice of calibration or hold-out sample, the process was repeated 1,000 times, each with randomly drawn choices of the two samples. The final success rate was the average across the 1,000 runs.

**Results:** The results of the comparison of the prediction success rates between the benchmark and CvS models are presented below in Tables S2 and S3, for the benchmark model without the average degree predictor and with it, respectively. It is possible to see from the tables, that the prediction success rate of the CvS model is highest for all cases and all prediction horizons. Especially interesting is that for the 10-year prediction horizon, the benchmark model loses all informative value, while the CvS model more or less retains its strength.

Table S2: Comparison of the prediction success rate between the benchmark and the CvS model for the model that does not include the average degree as the predictor. Each row presents a different prediction time horizon. The rate is measured in terms of percentage of success rate of predicting the correct tier, out of four tier sizes.

| Method | Benchmark success rates | Social circles theory success rates |
| --- | --- | --- |
| Next 120 days | 36.6% | 43.5% |
| Next year | 31.8% | 42.9% |
| Next 5 years | 25.1% | 39.8% |
| Next 10 years | 25.1% | 38.6% |

Table S3: Comparison of the prediction success rate between the benchmark and the CvS model for the model that includes the average degree as the predictor. Each row presents a different prediction time horizon. The rate is measured in terms of percentage of success rate of predicting the correct tier, out of four tier sizes.

| Method | Benchmark success rates | Social circles theory success rates |
| --- | --- | --- |
| Next 60 days | 44.2% | 45.2% |
| Next 120 days | 42.0% | 44.3% |
| Next year | 33.6% | 42.5% |
| Next 5 years | 25.1% | 39.2% |
| Next 10 years | 25.1% | 38.7% |